\documentclass[num-refs]{wiley-article}
\usepackage{nth}

\usepackage{enumitem}

\usepackage{CJKutf8}
\usepackage[utf8]{inputenc}
\usepackage{wasysym}
\usepackage{algorithm}
\usepackage[noend]{algpseudocode}
\usepackage{listings}
\usepackage{subcaption}

\lstdefinelanguage{JavaScript}{
  keywords={typeof, new, true, false, catch, function, return, null, catch, switch, var, if, in, while, do, else, case, break},
  keywordstyle=\color{blue}\bfseries,
  ndkeywords={class, export, boolean, throw, implements, import, this},
  ndkeywordstyle=\color{darkgray}\bfseries,
  identifierstyle=\color{black},
  sensitive=false,
  numberstyle=\color{red}\ttfamily,
  comment=[l]{//},
  morecomment=[s]{/*}{*/},
  commentstyle=\color{purple}\ttfamily,
  stringstyle=\color{blue}\ttfamily,
  morestring=[b]',
  morestring=[b]"
}

\usetikzlibrary{trees}
\lstdefinestyle{customjavascript}{%
   language=JavaScript,
   backgroundcolor=\color{white},
   extendedchars=true,
   basicstyle=\linespread{0.4}\footnotesize\ttfamily,
   showstringspaces=false,
   showspaces=false,
   numberstyle=\footnotesize,
   tabsize=2,
   breaklines=true,
   showtabs=false}

\lstdefinestyle{customc}{%
  belowcaptionskip=1\baselineskip,
  breaklines=true,
  xleftmargin=\parindent,
  language=C,
  showstringspaces=false,
  basicstyle=\small\ttfamily,
  keywordstyle=\bfseries\color{green!40!black},
  numberstyle=\tiny,
  commentstyle=\itshape\color{purple!40!black},
  identifierstyle=\bfseries\color{black},
  stringstyle=\color{orange},
   morekeywords={uint64_t,uint32_t,__m256i,__m128i,UINT64_C},
}

\newcommand{\breakabletexttt}[1]{%
  \begingroup
  \ttfamily
  \begingroup\lccode`~=`/\lowercase{\endgroup\def~}{/\discretionary{}{}{}}%
  \begingroup\lccode`~=`[\lowercase{\endgroup\def~}{[\discretionary{}{}{}}%
  \begingroup\lccode`~=`.\lowercase{\endgroup\def~}{.\discretionary{}{}{}}%
  \catcode`/=\active\catcode`[=\active\catcode`.=\active
  \scantokens{#1\noexpand}%
  \endgroup
}
\algnewcommand\algorithmicswitch{\textbf{switch}}
\algnewcommand\algorithmiccase{\textbf{case}}
\algnewcommand\algorithmicassert{\texttt{assert}}
\algnewcommand\Assert[1]{\State \algorithmicassert(#1)}%
\algdef{SE}[SWITCH]{Switch}{EndSwitch}[1]{\algorithmicswitch\ #1\ \algorithmicdo}{\algorithmicend\ \algorithmicswitch}%
\algdef{SE}[CASE]{Case}{EndCase}[1]{\algorithmiccase\ #1 \algorithmicdo}{\algorithmicend\ \algorithmiccase}%
\algtext*{EndSwitch}%
\algtext*{EndCase}%

\algnewcommand{\IIf}[1]{\State\algorithmicif\ #1\ \algorithmicthen}
\algnewcommand{\EndIIf}{\unskip\ \algorithmicend\ \algorithmicif}
\algnewcommand{\EElse}[1]{\State\algorithmicelse\ #1\ }
\algnewcommand{\EndEElse}{\unskip\ \algorithmicend\ \algorithmicelse}
\algnewcommand{\CCase}[1]{\State\algorithmiccase\ #1}
\algnewcommand{\EndCCase}{\unskip\ \algorithmicend\ \algorithmiccase}

\usepackage{algorithmicx}
\usepackage{tikzsymbols}
\usepackage[binary-units]{siunitx}
\usepackage{booktabs}
\usepackage{todonotes}
\usepackage{color}
\usepackage{multirow}
\usepackage{enumitem}
\definecolor{lightgray}{rgb}{.9,.9,.9}
\definecolor{darkgray}{rgb}{.4,.4,.4}
\definecolor{purple}{rgb}{0.65, 0.12, 0.82}
\usepackage{arydshln}
\usepackage{etoolbox}
\usepackage{xcolor}
\usepackage{listings}
\lstdefinestyle{customc}{%
  belowcaptionskip=1\baselineskip,
  breaklines=true,
  xleftmargin=\parindent,
  language=C,
  showstringspaces=false,
  basicstyle=\small\ttfamily,
  keywordstyle=\bfseries\color{green!40!black},
  numberstyle=\tiny,
  commentstyle=\itshape\color{purple!40!black},
  identifierstyle=\bfseries\color{black},
  stringstyle=\color{orange},
   morekeywords={uint64_t,uint32_t,__m256i,__m128i,simd8,uint8_t,UINT64_C},
}

\newcommand*{\restartrowcolors}{%
  \ifhmode\unskip\fi
  \vadjust{%
    \global\rownum=0 %
  }%
}

\papertype{Research Article}

\title{On-Demand JSON: A Better Way to Parse Documents?}

\author[1]{John Keiser}
\author[1\authfn{1}]{Daniel Lemire}

\affil[1]{DOT-Lab Research Center, Universit\'e du Qu\'ebec (TELUQ), Montreal, Quebec, H2S 3L5, Canada}

\corraddress{Daniel Lemire, DOT-Lab Research Center, Universit\'e du Qu\'ebec (TELUQ), Montreal, Quebec, H2S 3L5, Canada}
\corremail{lemire@gmail.com}

\fundinginfo{Natural Sciences and Engineering Research Council of Canada, Grant Number: RGPIN-2017-03910}

\runningauthor{John Keiser and Daniel Lemire}

\begin{document}

\maketitle

\begin{abstract}
JSON is a popular standard for data interchange on the Internet. Ingesting JSON documents can be a performance bottleneck. A popular parsing strategy consists in converting the input text into a tree-based data structure---sometimes called a Document Object Model or DOM. 
We designed and implemented a novel JSON parsing interface---called On-Demand---that appears to the programmer like a conventional DOM-based approach. However, the underlying implementation is a pointer iterating through the content, only materializing the results (objects, arrays, strings, numbers) lazily. 
On recent commodity processors, an implementation of our approach provides superior performance in multiple benchmarks. 
To ensure reproducibility, our work is freely available as open source software.
Several systems use On-Demand: e.g., Apache Doris, the Node.js JavaScript runtime, Milvus, and Velox.

\end{abstract}

\section{Introduction}
\label{sec:intro}
There are several text-based semi-structured document formats (e.g., HTML, XML, JSON). JSON is maybe the most popular format online for data interchange~\cite{rfc8259}. Several database systems such as CouchDB, RethinkDB, MongoDB, SimpleDB and JSON Tiles~\cite{10.1145/3448016.3452809} use JSON as their primary exchange format.

A JSON document must be stored in a valid Unicode (UTF-8) string. The JSON syntax is nearly a strict subset of the popular programming language JavaScript. It has four primitive types  (string, number, Boolean, null) and two composed types (arrays and objects). An object takes the form of a series of key-value pairs between braces where keys are strings and values can be primitive or composed types (e.g., \texttt{\{"name":"Jack","age":22\}}). An array is a list of comma-separated values (either primitive of composed) between brackets (e.g., \texttt{[1,"abc",null]}).  The JSON specification has six  \emph{structural characters} (`\texttt{[}', `\texttt{\{}', `\texttt{]}', `\texttt{\}}', `\texttt{:}', `\texttt{,}') to delimit the location and structure of objects and arrays.

Programmers rarely work directly on text-based semi-structured document formats: they prefer to work with a software interface---a \emph{parser}---providing automated validation and text-to-data conversion. There are two popular general-purpose parsing strategies~~\cite{10.1145/3503222.3507719}:
\begin{itemize}
\item Many parsers process the input document (e.g., JSON, HTML, XML) immediately into an in-memory data structure. Most web browsers use such an approach under the name Document Object Model (DOM): web pages (HTML) are loaded in memory into an ordered tree, accessible as a data structure from a programming language such as JavaScript~\cite{wood1998document}. Fig.~\ref{fig:jsonexample} illustrates how a JSON document might be viewed as an ordered tree. A DOM-like approach is convenient, but it requires that 
the original document be materialized in memory. If the programmer is only interested in a subset of the document, they must still construct the whole tree. Further, even when the programmer needs to ingest the whole document, they may need to copy the data into their own data structures and data types. Hence the materialization of the tree might be unnecessary and wasteful. 
\item There are also event-based or \emph{streaming} strategies. The document is processed from the beginning, and each newly encountered component is an event: e.g., the beginning of a string, the beginning of an array, the end of an array, etc. In some instances, the programmer may provide  functions corresponding to each event. 
E.g., a programmer might have a function that is triggered each time a string is encountered. 
One of the most popular families of such parsers might be Simple API for XML (SAX)~\cite{10.5555/515689}. Streaming approaches can be efficient: if the programmer only needs to capture part of the input document, they can ignore everything else. The programmer may also write the data to their own data structures directly, without the need to materialize a full tree in a temporary memory buffer. We may use these efficient strategies for specific tasks such as well-defined queries (e.g., JSONPath~\cite{10.1145/3297858.3304008}). We may also apply them to the  deserialisation of specific data structures: a popular framework that serves this purpose in Rust is called serde~\cite{scott2020using}. 
For general-purpose tasks, streaming strategies might be challenging to the programmer.
For example, the programmer may  need to write their own code to track their logical location within the document. Without help, the programmer may end up with streaming code that is not highly efficient.
\end{itemize}

\begin{figure*}\centering
\begin{tabular}{c|c}
\begin{minipage}[b]{0.5\columnwidth}
   \lstset{escapechar=@,style=customjavascript}
\begin{lstlisting}
{
	"Width": 800,
	"Height": 600,
	"Title": "View from my room",
	"Url": "http://ex.com/img.png",
	"Private": false,
	"Thumbnail": {
		"Url": "http://ex.com/th.png",
		"Height": 125,
		"Width": 100
	},
	"array": [
		116,
		943,
		234
	],
	"Owner": null
}
\end{lstlisting}
\end{minipage}
&
\tikzstyle{every node}=[draw=black,anchor=west,fill=blue!5]
\tikzstyle{selected}=[draw=red,fill=red!30]
\tikzstyle{optional}=[dashed,fill=gray!50]
\begin{tikzpicture}[%
  scale=0.5, transform shape,
  grow via three points={one child at (0.5,-0.7) and
  two children at (0.5,-0.7) and (0.5,-1.4)},
  edge from parent path={(\tikzparentnode.south) |- (\tikzchildnode.west)}]
  \node {root}
    child { node {"Width": 800}}		
    child { node {"Height": 600}}
    child { node {"Title": "View from my room"}}
    child { node {"Url": "http://ex.com/img.png"}}
    child { node {"Private": false}}
    child { node {"Thumbnail"}
      child { node {"Url": "http://ex.com/th.png"}}
      child { node {"Height": 125}}
      child { node {"Width": 100}}
    }
    child [missing] {}				
    child [missing] {}				
    child [missing] {}		
    child { node {"array"}
      child { node {116}}
      child { node {943}}
      child { node {234}}
    }
    child [missing] {}				
    child [missing] {}				
    child [missing] {}	
    child { node {"Owner": null}};
\end{tikzpicture}
\end{tabular}
\caption{\label{fig:jsonexample} JSON example with corresponding tree form}
\end{figure*}

It is possible to  parse JSON at high speed with DOM-like approaches. The simdjson DOM parser~\cite{langdale2019parsing} can construct an ordered tree at a speed of over \SI{2}{\gibi\byte\per\second} on realistic inputs. However, we
might be able to go even faster if we skip the in-memory construction of the ordered tree. The conventional approach to avoid the materialization of the document as an in-memory data structure is to adopt a streaming strategy. However, even when it would provide superior performance (e.g., $2\times$ or $3 \times$ faster), we believe that many programmers would still prefer the DOM-like approach for convenience. To illustrate, consider the source code needed to load coordinates stored as JSON objects (e.g., \{"x":1,"y":2,"z":3\}) using the standard (DOM-like) interface of the popular JSON for Modern C++ library:\footnote{\url{https://github.com/nlohmann/json}}
\begin{lstlisting}    
auto root = nlohmann::json::parse(json.data(), json.data() + json.size());
for (auto point : root["coordinates"]) {
  result.emplace_back(json_benchmark::point{point["x"], point["y"], point["z"]});
}
\end{lstlisting}
We implemented the same solution using the streaming interface from the same library (JSON for Modern C++), see Fig.~\ref{fig:jsonstreamexample}. For simplicity, we omitted boilerplate code which handles unexpected events. Even so, the streaming interface is relatively complicated, significantly longer (4 lines vs. 40), and requires much effort from the programmer compared to the standard imperative DOM approach. The code might be also more difficult to debug.

And this is a simple example: more complex processing would unavoidably require more code. It might be possible to improve our solution, but an event-based programming approach unavoidably requires the programmer to provide its own approach to track the position in the document and its current state.

\begin{figure*}\centering
\lstset{language=C++, xleftmargin=.2\textwidth, xrightmargin=.2\textwidth,basewidth=0.5em, showspaces=false, showtabs=false}
\begin{lstlisting}    
struct Handler : json::json_sax_t {
  size_t k{0}; double buffer[3]; std::vector<point> &result;
  Handler(std::vector<point> &r) : result(r) {}

  bool key(string_t &val) override {
    switch (val[0]) {
    case 'x': k = 0; break;
    case 'y': k = 1; break;
    case 'z': k = 2; break;
    }
    return true;
  }
  bool number_float(number_float_t val, const string_t &s) override {
    buffer[k] = val;
    if (k == 2) {
      result.emplace_back(
          json_benchmark::point{buffer[0], buffer[1], buffer[2]});
      k = 0;
    }
    return true;
  }
  bool number_unsigned(number_unsigned_t val) override {
    buffer[k] = double(val);
    if (k == 2) {
      result.emplace_back(
          json_benchmark::point{buffer[0], buffer[1], buffer[2]});
      k = 0;
    }
    return true;
  }

  bool parse_error(std::size_t position, const std::string &last_token,
                   const json::exception &ex) override {
    return false;
  }
}; // Handler


// ...
Handler handler(result);
json::sax_parse(json.data(), &handler);
\end{lstlisting}
\caption{\label{fig:jsonstreamexample} Example of a streaming approach to parse coordinate data in JSON using JSON for Modern C++}

\end{figure*}

For high-efficiency and convenience,  
we want to offer to the programmer a lazily evaluated tree~\cite{henderson1976lazy}: if the programmer only needs one number out of a larger document, we  allow the programmer to navigate to this one number using as little effort as possible, and to only materialize this one number. When the input document is an array (e.g., \texttt{[1,2,3]}), we propose to offer to the programmer what appears to be an array data structure, but is just an iterator over the values in the array. For objects (e.g., \texttt{\{"one":1, "two":2, "three":3\}} in JSON), we iterate over key-value pairs. The parsing of the values is delayed until it is needed: the programmer may choose to skip unneeded values. Furthermore, we seek to provide as much flexibility to the programmer as possible. A given value appearing in JSON as a number may be parsed by the programmer as an integer, a floating-point number, or  a  string; a string (e.g., \texttt{"3.1416"}) can be parsed as number; and so forth.  We call this lazy parsing strategy \emph{On-Demand}. We refer to  On-Demand as a \emph{front-end}: it is a programming interface for the programmer that serves to abstract much of the complexity necessary for correct and efficient parsing.

An open-source implementation of our proposal (simdjson On-Demand) provides
high speed. 
A long-standing benchmark of several JSON parsers\footnote{\url{https://github.com/kostya/benchmarks}} (henceforth Kostya) starts
with a \num{10000}-long array of coordinates in JSON (e.g., \breakabletexttt{\{"coordinates":[\{"x":2.0,"y":0.5,"z":0.25\}, \ldots]\}}) and requires that 
the parser sums all `x` values, all `y` values and all `z` values. 
The On-Demand source code specific to this benchmark is provided in Fig.~\ref{fig:kostyaondemand}.
The results are updated regularly, but the On-Demand approach is often 
ranked in first position among $\approx 75$~competitors.
We present some of the results in Table~\ref{tab:kostya} for July~2022.
The nearest competitor which does not sacrifice floating-point accuracy
is Serde in the Rust programmming language. 
A highly efficient C++ approach (DAW JSON Link) is practically
on par with On-Demand in performance and it offers a functionality similar to
Serde, with other specialized compile-time constructions, although without  exact number parsing~\cite{langdale2019parsing,lemire2021number}.

We believe that the Kostya benchmark illustrates 
that an On-Demand front-end may achieve high performance with a software interface that is both expressive---our code is generic---and accessible: our users' code resembles that of a DOM interface.

\begin{table}
    \centering
    \begin{tabular}{ccc}
\toprule
Library and  Language  &                Time (s) &  Energy Usage (J)  \\\midrule
C++/g++ (simdjson On-Demand) &  0.067   & 2.8\\
 Rust (Serde) &  0.11 &  5.6  \\
 C++/g++ (RapidJSON) &  0.25 & 6.9 \\
 C++/g++ (Boost.JSON) &  0.40 & 16 \\
 Go& 	0.86 & 36 \\
C++/clang++ (Nlohmann) &  1.3 & 53 \\
Python &	1.5  & 60 \\
\bottomrule
    \end{tabular}
\caption{\label{tab:kostya}Sample of the kostya JSON benchmark results (July 2022), after excluding techniques that fail to provide exact number parsing. The benchmark is reported
to run on an Intel Xeon E-2324G processor (Rocket Lake), using a Debian distribution with GCC 12.1, Go 1.18.2, Rust 1.61 and Python 3.10.4.
The test file is \SI{115}{\giga\byte}, the timing excludes the time requires to load the file from disk. Memory usage is computed with from Intel performance counters.}

\end{table}

\begin{figure}    \centering
\lstset{language=C++, xleftmargin=.2\textwidth, xrightmargin=.2\textwidth,basewidth=0.5em}

\begin{lstlisting}
  auto doc = parser.iterate(json);
  for (object point_object : doc["coordinates"]) {
    x += double(point_object["x"]);
    y += double(point_object["y"]);
    z += double(point_object["z"]);
  }
\end{lstlisting}
\caption{\label{fig:kostyaondemand}On-Demand C++ source code used by the kostya JSON benchmark}

\end{figure}

\section{Related Work}

Our work builds on simdjson~\cite{langdale2019parsing}, specifically on its indexing stage. The simdjson parser was the first standard-compliant JSON parser to process gigabytes of data per second on a single core using commodity processors. Given an input JSON document, 
the simdjson parser  identifies  the starting location of all JSON nodes (e.g., numbers, strings, null, true, false, arrays, objects)
as well as all JSON structural characters (`\texttt{[}', `\texttt{\{}', `\texttt{]}', `\texttt{\}}', `\texttt{:}', `\texttt{,}'). 
The parser stores these locations as integer indexes in a separate array. For example, given the array \texttt{\{"abc":2000\}}, we might have the indexes 0, 1, 6, 7, 11.
During this indexing stage, we must distinguish the characters that are between quotes, and thus inside a
string value, from other characters. For example, the JSON document \texttt{"[1,2]\textbackslash{}""} is a single string and we only need to
locate the first quote character.
During the indexing phase, the simdjson parser also checks the Unicode encoding using an efficient SIMD-based algorithm~\cite{keiser2021validating}. It also checks other rules such that we only find the ASCII space, the horizontal tab, the line feed, the carriage return  and the JSON structural characters  (`\texttt{[}', `\texttt{\{}', `\texttt{]}', `\texttt{\}}', `\texttt{:}', `\texttt{,}') outside strings.
The entirety of the indexing is done in a bit-parallel manner, and leveraging SIMD instructions when available.
It reads blocks of 64~bytes one after the other. From the 64~bytes, it generates a 64-bit word which acts as a bitset: the location of each 1-bit indicates the potential location of a JSON structural character or the start of a value. 
From this 64-bit word, we extract index locations to an array of integers, each 64-byte block generating between zero to 32~index values. 
The software relies on runtime dispatching, first identifying the features supported by the CPU and then calling on the appropriate
function.
The simdjson library is used by popular database systems such as ClickHouse and StarRocks.

Though the simdjson parser provides state-of-the-art speed on single-core commodity processors,  there are faster JSON parsers on more specialized hardware. Using field-programmable gate arrays (FPGA), Dann et al.~\cite{10.1145/3533737.3535094} designed PipeJSON, the first standard-compliant JSON parser to process tens of gigabytes of data per second. It consists in a JSON parser prototype with a simdjson-compatible interface, and it can serve as a drop-in replacement for the simdjson parser.
Hahn et al.~\cite{hahn2022raw} present a fast FPGA JSON filtering parser than can locate relevant components with high accuracy and high speed. Peltenburg et al.~\cite{peltenburg2021tens} convert JSON to the Arrow format at the rate of tens of gigabytes per second on an FPGA.
Stehle and Jacobsen similarly achieve high speeds, using a graphical processing unit (GPU)~\cite{10.14778/3377369.3377372}. It is also possible to accelerate JSON parsing with multicore parallelism~\cite{10.14778/3436905.3436926,pavlopoulou2018parallel}.

One of the advantages of our On-Demand approach is that the programmer can quickly skip irrelevant JSON data.
Consuming JSON faster by skipping irrelevant sections is a common strategy. Alagiannis et al.~\cite{Alagiannis:2012:NAA:2367502.2367543} query  JSON without  loading it in their database---parsing only what is necessary. Bonetta and Brantner use  just-in-time (JIT) compilation and  selective data access~\cite{Bonetta:2017:FFJ:3137765.3137782}. 
Jiang and Zhao~\cite{10.1145/3503222.3507719} implemented a JSON streaming framework (JSONSki)
to fast-forward over different cases of irrelevant substructures.
 Li et al.\ present their fast parser, Mison which can jump directly to a queried field without parsing intermediate content~\cite{Li:2017:MFJ:3115404.3115416}. Mison uses SIMD instructions to quickly identify some structural characters but otherwise works by processing bit-vectors in general purpose registers with branching loops. Mison preceded simdjson and it is broadly similar to simdjson's indexing.
FishStore~\cite{Xie:2019:FFI:3299869.3319896} parses  JSON data and selects subsets of interest, storing the result in a fast key-value store~\cite{Chandramouli:2018:FCK:3183713.3196898}. Though built originally on Mison, 
FishStore has adopted simdjson\footnote{\url{https://github.com/microsoft/FishStore}}. Palkar et al. present Sparser: it filters quickly an unprocessed document to find mostly just the relevant information~\cite{palkar2018filter}, and then relies on a  parser.


\section{On-Demand Front-End}

Conceptually, parsing with On-Demand begins with the creation of a parser instance (\texttt{ondemand::parser}). The parser is responsible for memory allocation and it can be reused from document to document. Given a parser instance, the user calls the \texttt{iterate} method on a string input which returns a document instance (\texttt{ondemand::document}). The method calls the
simdjson indexing routine, and writes an index in the parser instance.
Afterward, the front-end consists essentially in a pointer over the document which we call internally
a \texttt{json\_iterator}: the iterator resides within the document instance.

We typically traverse the document
sequentially from the beginning to the end, always pointing at one pseudo-structural character at a time. The
\texttt{json\_iterator} has also a reference to a pre-allocated string buffer---owned by
the parser instance---where we may decode JSON strings: strings in JSON may contain escaped characters 
(e.g., \texttt{\textbackslash{}n} or \texttt{U0030}) and we provide the user with an unescaped version stored in our own buffer.

During the processing of a valid JSON document, we point at a structural character 
(`\texttt{[}', `\texttt{\{}', `\texttt{]}', `\texttt{\}}', `\texttt{:}', `\texttt{,}') or at the beginning
of a value:
\begin{itemize}
\item A quote character (`\texttt{"}') marking the beginning of either a string value or an object key.
\item A digit or minus character (`\texttt{-}', `\texttt{0}', \ldots, `\texttt{9}') marking the beginning of number.
\item The letters `\texttt{n}', `\texttt{t}' or `\texttt{f}' for the start of a \texttt{null}, \texttt{true}, or \texttt{false} token. 
\end{itemize}
However not all of these states are accessible to the programmer: in the normal course
of operation, the code points at the beginning
of an array (`\texttt{[}'), at the beginning of an object (`\texttt{\{}') or at the beginning of a value (number, string, Boolean, null). 

The On-Demand front-end validates the content that it parses. We may choose to skip a value (e.g., a string
or a number) and
in which case the content of the value is not fully validated. For example, the JSON string
\texttt{[1,1b]} is invalid---because \texttt{1b} is not a valid number string---but if it is not accessed (i.e., it is skipped),
then no error may be reported. 

To help us keep track of the structure of the JSON document, 
our pointer instance (\texttt{json\_iterator}) contains a depth counter. When we enter
an array or an object, the depth is incremented by one, and decremented by one when we exit the object or the array.
In case of an error or when we have consumed the whole document, the depth is set to zero.
Table~\ref{table:depthmodel} illustrates the depth model for the document from Fig.~\ref{fig:jsonexample}.

From the point of view of the programmer, we create arrays and objects 
(\texttt{ondemand::array} and \texttt{ondemand::object}). 
Yet these instances are mere thin wrappers and the creation of the instance translates into a check for the presence of the 
right structural characters (`\texttt{[}', `\texttt{\{}', `\texttt{]}', `\texttt{\}}'). They record their starting position and their document depth. 
They provide
access to corresponding iterators over the elements
of the arrays or the fields of the object.

When reading an object's key-value
pair, the key's location is accessed and the pointer is moved to the associated value. We return to the user a \texttt{field} structure 
containing a pointer to the key that was just read, and a reference to the value.
We may also seek a key within an object: when doing so, we scan for the presence of string instances (recognized by a quote
character `\texttt{"}') preceded by a colon character (`\texttt{:}')
at the depth of the current object. The programmer might seek different key values 
as in Fig.~\ref{fig:kostyaondemand}: each time we advance through the content, with a possible restart from the beginning if we
arrive at the end of the object. We know that a key cannot be found if we have scanned through the object entirely, with a possible
restart from the beginning, arriving back to our starting point. We always scan for keys from our current position, instead 
of systematically starting from the beginning.
If the key is not found then we are left pointing at the either the start of the object or at a comma. 
We are at the depth of the object itself.
When searching for a key, if the key is found then we are left pointing at the colon, and we are one depth below the object itself.
We then immediately advance to the value.

As much as possible, we rely on \texttt{std::string\_view} instances. Unlike the conventional C++ \texttt{std::string}
which allocate their own memory, the \texttt{std::string\_view} instances
are thin wrappers around a character pointer. Thus when we return a string value, we do not need to allocate 
a new buffer. 
Given a JSON value (string, number, Boolean or null), the \texttt{raw\_json\_token()} method provides
a direct (unprocessed) \texttt{std::string\_view} instance mapped to the original JSON string. Because
the document is indexed, such a method can be implemented using few instructions. Given an array or an object, a programmer can have direct access to the corresponding memory region 
in the original document: the \texttt{raw\_json()} method returns a \texttt{std::string\_view} instance
representing the unprocessed string representation of the array or object.

\begin{table}\centering
\begin{tabular}{llc}
\toprule
data type &  pointed text (prefix) & depth \\ \midrule
\texttt{   +document         } & \texttt{ \{  "Width": 800,  "Height": 60} &    1\\
\texttt{   +object           } & \texttt{ \{  "Width": 800,  "Height": 60} &     1 \\
\texttt{     double          } & \texttt{ 800,  "Height": 600,  "Title":} &     2 \\
\texttt{     double          } & \texttt{ 600,  "Title": "View from my r} &     2 \\
\texttt{     string          } & \texttt{ "View from my room",  "Url": "} &     2 \\
\texttt{     string          } & \texttt{ "http://ex.com/img.png",  "Pri} &     2 \\
\texttt{     bool            } & \texttt{ false,  "Thumbnail": \{   "Url"} &     2 \\
\texttt{     bool            } & \texttt{ false,  "Thumbnail": \{   "Url"} &     2 \\
\texttt{     +object         } & \texttt{ \{   "Url": "http://ex.com/th.p} &     2 \\
\texttt{       string        } & \texttt{ "http://ex.com/th.png",   "Hei} &     3 \\
\texttt{       double        } & \texttt{ 125,   "Width": 100  \},  "arra} &     3 \\
\texttt{       double        } & \texttt{ 100  \},  "array": [   116,   9} &     3 \\
\texttt{     -object         } & \texttt{ \},  "array": [   116,   943,  } &     2 \\
\texttt{     +array          } & \texttt{ [   116,   943,   234  ],  "Ow} &     2 \\
\texttt{       double        } & \texttt{ 116,   943,   234  ],  "Owner"} &     3 \\
\texttt{       double        } & \texttt{ 943,   234  ],  "Owner": null } &     3 \\
\texttt{       double        } & \texttt{ 234  ],  "Owner": null \}      } &     3 \\
\texttt{     -array          } & \texttt{ ],  "Owner": null \}           } &     2 \\
\texttt{     skip            } & \texttt{ null \}                        } &     2 \\
\texttt{   -object           } & \texttt{ \}                             } &     1 \\
\end{tabular}
\caption{\label{table:depthmodel}Representation of the state of our On-Demand pointer as we move through the 
document. The `+' and `-' prefix on the data types indicate the beginning and end of
arrays and objects respectively.}
\end{table}

Most functions may return an error. 
To provide flexible error handling, our functions
often return a value of a type corresponding to the template \texttt{simdjson\_result}. This template contains a pair of values: an error code and the actual value. Hence, a \texttt{get\_double()} method might return a structure containing an error code and a 64-bit floating-point value of the type \texttt{double}. A \texttt{simdjson\_result} instance throws an exception when we attempt to cast it to its value type (e.g.,  \texttt{double}) if the error condition indicates an error. 
For users who prefer to avoid exception handling, we make it possible to recover the value from a \texttt{simdjson\_result} instance after checking the error condition.
Most of our methods (\texttt{get\_array()}, 
\texttt{get\_object()}, 
\texttt{get\_bool()}, \ldots) return  \texttt{simdjson\_result} instances. 
For convenience, we can cast values directly to a given type (e.g., \texttt{double}) without explicitly creating a \texttt{simdjson\_result} instance.

Fig.~\ref{fig:ondemandexample} illustrates
how the On-Demand front-end might work with a non-trivial example. Given a JSON document made of an array of objects, we initiate a loop, accessing each element of the array as an object (\texttt{car}). Given such an object, we may query some fields (e.g., \texttt{model}) but not others (e.g., \texttt{make}): unqueried fields are skipped as much as possible. We may cast a value to a string, to an integer or to a floating-point value. In this particular example, if the value does not match the desired type, then a C++ exception would be thrown.
We also support arbitrary schema through the \texttt{type()} method: it indicates whether the current node is potentially a number, a string, a Boolean, an array, an object, etc.

\begin{figure}    \centering
\lstset{language=C++, xleftmargin=.2\textwidth, xrightmargin=.2\textwidth,basewidth=0.5em, showspaces=false, showtabs=false}

\begin{lstlisting}
ondemand::parser parser;
auto cars_json = R"([
{"make":"Toyota","model":"Camry","year":2018,
"tire_pressure":[40.1,39.9,37.7,40.4]},
{"make":"Kia","model":"Soul","year":2012,
"tire_pressure":[30.1,31.0,28.6,28.7]},
{"make":"Toyota","model":"Tercel","year":1999,
"tire_pressure":[29.8,30.0,30.2,30.5]}
])"_padded;
// Iterating through an array of objects
for (ondemand::object car : parser.iterate(cars_json)) {
  // Accessing a field by name
  cout << "Model: " << std::string_view(car["model"]) << endl;
  // Casting a JSON element to an integer
  uint64_t year = car["year"];
  cout << "- This car is " << 2020 - year << "years old." << endl;
  // Iterating through an array of numbers
  double total_tire_pressure = 0;
  for (double tire_pressure : car["tire_pressure"]) {
    total_tire_pressure += tire_pressure;
  }
  cout << "- Average tire pressure: " << (total_tire_pressure / 4) << endl;
}
\end{lstlisting}
\caption{\label{fig:ondemandexample}On-Demand C++ example: it illustrates how we access key-value elements, handle integers and strings.}
\end{figure}



The simdjson library offers both a conventional DOM implementation and an On-Demand approach. The On-Demand approach requires less memory in general. While both strategies materialize the index, which uses one word (four bytes) per pseudo-structural character, the On-Demand approach may not require any other intermediate memory.

In the DOM implementation of simdjson, a tree-like data structure is written during parsing, which is later accessed by the programmer. This data structure simply does not exist in the On-Demand approach.

We designed the simdjson library so that memory usage is owned by the parser instance. We expect users facing memory-intensive conditions to reuse the same parser repeatedly, which may eliminate the need to allocate new memory as long as the size of the JSON documents is bounded.

The On-Demand design has some known limitations. On-Demand operates with a single pointer within the document. This means that if a user needs to process multiple components of the documents simultaneously, they must either materialize the data into their own data structure or rewind the iterator to earlier sections of the document. Rewinding the iterator may lead to reparsing the same JSON data multiple times, resulting in a performance penalty.

To illustrate this effect, consider a user who acquires an object instance and repeatedly queries a key value (e.g., \texttt{object["mykey"]}). Doing so might require scanning the JSON data of the object multiple times for the same key. In this scenario, the programmer should attempt to materialize the value once because On-Demand does not currently support caching.

 An  anti-pattern for On-Demand would be this code sequence:
\begin{lstlisting}
 std::string_view make = o["data"]["make"];
 std::string_view model = o["data"]["model"];
 std::string_view year = o["data"]["year"];
\end{lstlisting}
A more performant solution would search for the key \texttt{data} just once:
\begin{lstlisting}
ondemand::object data = o["data"];
std::string_view model = data["model"];
std::string_view year = data["year"];
std::string_view rating = data["rating"];
\end{lstlisting}
In general, On-Demand provide an interface that appears like a DOM, but it has different performance characteristics  and it requires some care to provide optimal performance. We therefore encourage the programmers to run their performance-sensitive code with logging enabled, so that the programmer can see the various tasks that the front-end has to carry.

The On-Demand provides complete error handling, and it is even possible to know where (in the JSON document) the error occurred, but it can be surprising for some users that an error is triggered after they have seemingly loaded the JSON document. Indeed, the On-Demand front-end generally accepts invalid documents, and it only validates the components that the user parses, as they are being parsed. It requires slightly more care on the part of the programmer because the errors can happen at more locations in their code.

\section{Experiments}

We benchmark our On-Demand front-end on a recent
Intel architecture. 
We use a server configured with Rocky~Linux~9.
The server has two 32-core  Intel Xeon Gold 6338  (Ice Lake) processors having
\SI{48}{\mebi\byte} of L3 memory: the 64~cores have \SI{48}{\kilo\byte} of L1~data cache memory and 
\SI{1.24}{\mebi\byte} of L2 cache memory.
They are rated with a base clock frequency of \SI{2.0}{\giga\hertz} and
a maximal frequency of \SI{3.2}{\giga\hertz}.
The server has 
\SI{376}{\gibi\byte}
 of main memory (DDR4, \SI{3200}{\mega\hertz}).
The benchmarks are single-threaded and we exclude disk and network accesses
from our tests.

The software is written in C++ and compiled with GCC~12 using the default
CMake setting for a release build: \texttt{-O3 -DNDEBUG}. 
We do not compile for a specific processor architecture.  The benchmarks
are written using the Google Benchmark (v1.6.0) framework.\footnote{\url{https://github.com/google/benchmark}} Our benchmarking code is \emph{instrumented}: we use the performance counters of the processors to record the number instructions retired, the number of cycles and the number of mispredicted branches. 
It allows us to compute the actual average processor frequency:
we verify that it can reach \SI{3.2}{\giga\hertz} as expected.
The number of instructions \emph{retired} includes only the instructions that have been fully executed and excludes instructions issued solely due to \emph{speculative execution}~\cite{guide2016intel,marques2021mansard,gravelle2022performance}.

To benchmark and validate our work, we need various tasks.
Our goal is to 
compare On-Demand with existing generic JSON parsers.
We choose to compare against the following state-of-the-art 
C++ standard-compliant parsers. 
\begin{itemize}
    \item The yyjson parser is a C library that was released shortly after the first release of the simdjson library. According to the author (private communication), 
    the goal of the yyjson was to get as close as possible to simdjson, while using
    only generic C code.  We use the version 0.5.1, released on June 17, 2022. 
    \item In earlier work, Langdale and Lemire selected RapidJSON and sajson as references. In 2018, RapidJSON was described as the fastest traditional state-machine-based parser available~\cite{palkar2018filter}. Though they both have similar performance, sajson only does partial UTF-8 validation. Given the introduction of yyjson, we keep only RapidJSON (version 1.1.0, released in September 2018).
    \item We also include JSON for Modern C++ (version 3.10.5, from January 2022)---it is also known as Nlohmann/json. It is a popular parser.
\end{itemize}
Though there are many other libraries, in different programming languages, we believe that these three choices represent a good sample of competitive solutions. To test On-Demand, we use the simdjson library version~3.2.3 (released in August 2023\footnote{\url{https://github.com/simdjson/simdjson}}).

We use the JSON input as a database, and we seek to resolve some queries. We load all JSON documents in memory prior to running the benchmark; we omit disk and network accesses. Our main test document is the \texttt{twitter.json}~file~\cite{langdale2019parsing}: it is the result of a search for the character one in Japanese and Chinese using the Twitter API. It contains \num{631515}~bytes, \num{2108}~integers, \num{18099}~strings, \num{1264}~objects and \num{1050}~arrays.
We picked the following tasks. 

\begin{itemize}
    \item \emph{json2msgpack}: We must validate and convert   the \texttt{twitter.json} file into a binary MessagePack (MsgPack for short) equivalent. MessagePack is a binary format comparable to JSON~\cite{ching2018rcppmsgpack}. We use a simple conversion: all numbers are mapped to double types, all strings are prefixed by a 32-bit counter, and so forth. All our implementations rely on a recursive function call where the type of the current document node is queried and used as part of a switch/case query. When an array or an object is encountered, its content is processed recursively. It is an instance where the entire content of the file is consumed.
    \item \emph{partial tweets}: The \texttt{twitter.json} file consists of an object containing a field with the key \texttt{statuses} having as a value an array. Each element of the array is a tweet as a JSON object. 
    We seek to extract from each tweet the fields \texttt{created\_at}, \texttt{id}, \texttt{text}, \texttt{in\_reply\_to\_status\_id}, \texttt{retweet\_count}, \texttt{favorite\_count} and from a subobject associated with the key \texttt{user}, we want to acquire the  user's \texttt{id} and the user's \texttt{screen\_name}. We expect that such partial extraction of the content of a JSON document is common in practice. 

    \item \emph{distinct user}: In the \texttt{twitter.json} file, we seek to find all user identifiers (\texttt{id}) as 64-bit identifiers and store them in an array implemented as a \texttt{std::vector<uint64\_t>} instance. They are found as part of tweets (\texttt{user/id}) or within retweet metadata (keys \texttt{retweeted\_status/user/id}).
    
    \item \emph{find tweet}: We seek the textual content (\texttt{text}) of the tweet having identifier \texttt{505874901689851904}. There is only one such tweet.
    
    \item \emph{top tweet}: We scan all tweets and report the most retweeted one (according to \texttt{retweet\_count}), returning the \texttt{screen\_name} of the author and the textual content (\texttt{text}).
\end{itemize}
We also extended our benchmarks with two synthetic data sources:
\begin{itemize}

    \item \emph{kostya}: We reproduce the Kostya benchmark (see \S~\ref{sec:intro} and Fig.~\ref{fig:kostyaondemand}),  with less overhead and in a more controlled setting (Google~Benchmark). We process \num{524288}~triples stored in memory. Whereas the original Kostya benchmark computes the sums of the \texttt{x}, \texttt{y}, and \texttt{z} values, we store the parsed triples in a dynamic array (\texttt{std::vector}). 

    \item \emph{large random}: We proceed with a task similar to the Kostya benchmark except that instead of summing up the values, we construction three-value vectors and we append them to an array. We create, in memory, a temporary JSON document made of \num{1000000} triples of floating-point values within objects (e.g., \texttt{{ "x":0.14323412321,  "y":0.931321111, "z":0.0111141232312}}).
    
\end{itemize}

On a recent Linux system with the prerequisite (CMake version 3.15 or better and GCC 12), we expect our benchmarks to be reproducible using a few command lines, e.g.:
\begin{verbatim}
git clone https://github.com/simdjson/simdjson.git
cd simdjson
cmake -B build -DSIMDJSON_DEVELOPER_MODE=ON
cmake --build build -j
./build/benchmark/bench_ondemand
\end{verbatim}

Table~\ref{table:proc} presents the processing speed for various tasks. We
define the speed as the size of the input divided by the estimated elapsed time. Performance timings are heavily skewed toward the minimum; they do not follow a normal distribution~\cite{hoefler2015scientific}.  Accordingly, we present the best processing speed (out of all runs) and the average processing speed. Except for the \emph{json2msgpack} task, the On-Demand front-end is more than twice as fast as any competitor, except maybe for the DOM version of simdjson. Considering the geometric mean of the best results, we find that On-Demand is 70\% faster than the conventional (DOM-based) simdjson, it is over 2.5~times faster than yyjson, and nearly eight times faster than RapidJSON.  On-Demand is nearly 50~times faster than JSON for Modern~C++.

\begin{table}\small
\caption{\label{table:proc}Best and average processing speed in \si{\gibi\byte\per\second} for various tasks and implementations.}
\begin{tabular}{l|ccccc}\toprule
task                &  simdjson (On-Demand) &    simdjson (DOM) & yyjson & RapidJSON &  JSON for Modern C++ \\\midrule
json2msgpack& 2.3--2.5 & 1.7--1.8 & 0.75--1.3 & 0.38--0.43 & 0.023--0.025 \\
partial tweets& 4.8--5.2 & 2.9--3.1 & 0.96--2.0 & 0.38--0.48 & 0.090--0.10 \\
distinct user & 5.0--5.4 & 2.9--3.2 & 0.96--2.1 & 0.38--0.50 & 0.098--0.11 \\
find tweet& 8.0--8.7 & 3.1--3.3 & 0.97--2.1 & 0.38--0.49 & 0.12--0.13 \\
top tweet& 4.9--5.3 & 3.0--3.2 & 0.97--2.0 & 0.38--0.55 & 0.080--0.086 \\
kostya& 2.5--2.7 & 1.5--1.6 & 0.97--1.0 & 0.54--0.57 & 0.094--0.099 \\
large random& 0.87--0.91 & 0.48--0.51 & 0.45--0.47 & 0.24--0.25 & 0.043--0.045 \\\midrule
geometric mean& 3.3--3.6 & 1.9--2.1 & 0.84--1.4 & 0.37--0.46 & 0.069--0.075 \\

\bottomrule
\end{tabular}
\end{table}

Table~\ref{table:ins} presents the number of CPU instructions retired by input byte. The number of instructions required is stable in our tests (within 1\%). We find that the On-Demand front-end requires far fewer instructions than the competitors. 
It is indicative of a low number of mispredicted branches and of limited data dependencies.
When considering the geometric mean, On-Demand uses 60\% of the instructions of the conventional (DOM-based) simdjson, half of yyjson's instructions, and nearly eight times fewer instructions than RapidJSON. Compared to JSON for Modern C++, On-Demand requires over 40~times fewer instructions. In general, On-Demand is even faster than the numbers suggest, since it typically achieves a high number of instructions retired per unit of time. This is indicative of a low number of mispredicted branches and limited data dependencies.

\begin{figure}\centering
 \begin{subfigure}[h]{0.49\textwidth}
 \includegraphics[width=0.99\textwidth]{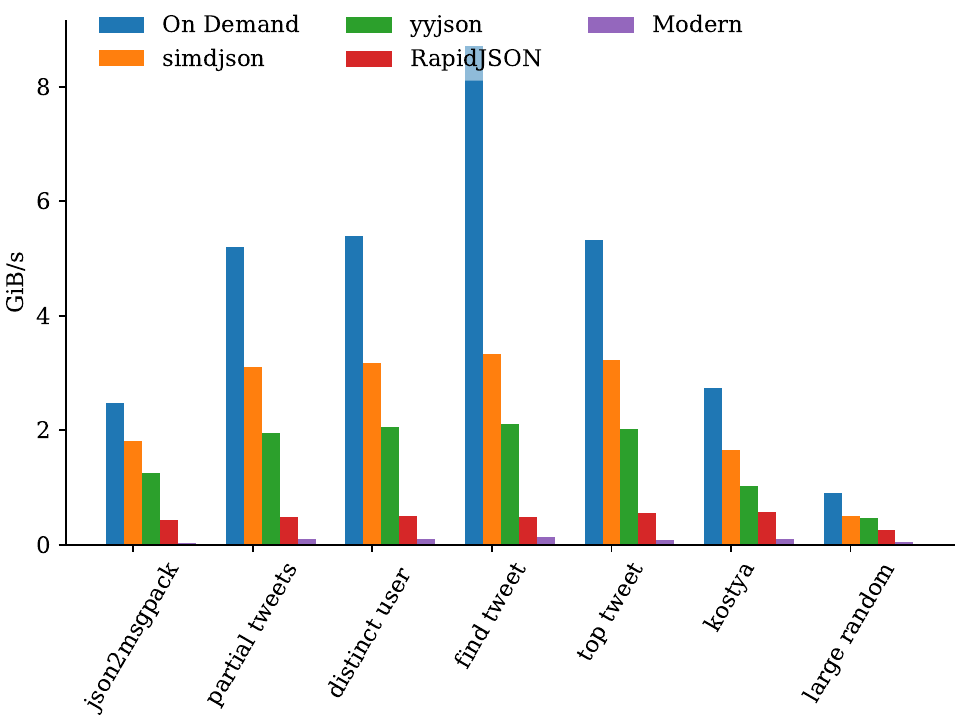}
\caption{Best speed (gigabytes per second)} \end{subfigure}
 \begin{subfigure}[h]{0.49\textwidth}
 \includegraphics[width=0.99\textwidth]{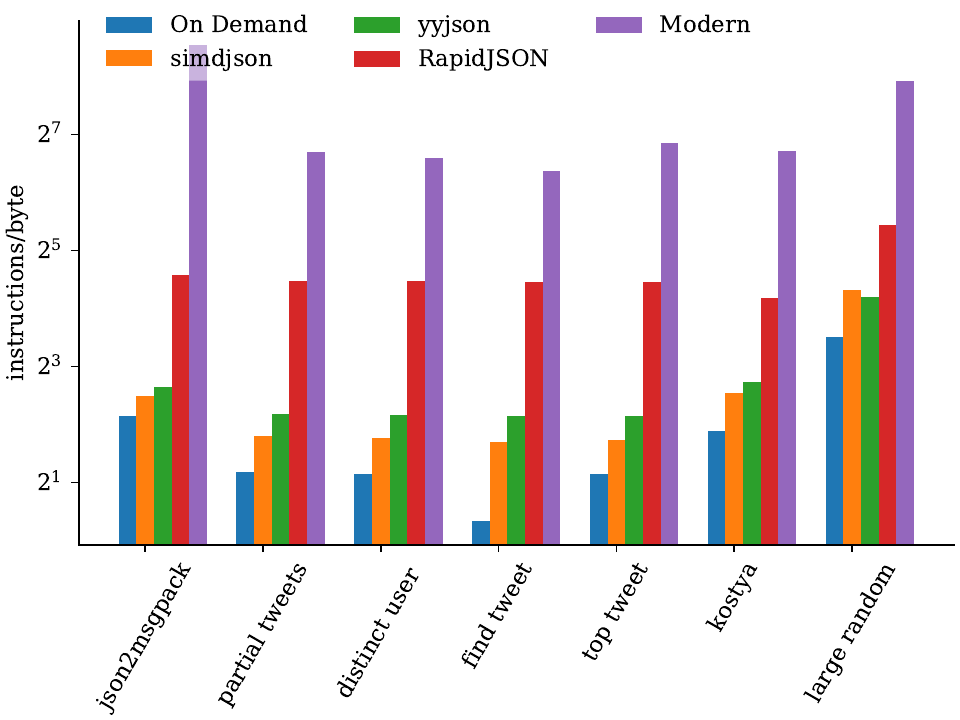}
\caption{instructions per byte} \end{subfigure}
\caption{Speed and number of instructions for various benchmarks. 
\label{fig:plots}}
\end{figure}

Fig.~\ref{fig:plots} presents the speed and number of instructions in graphical form. JSON for Modern~C++ is much slower than the alternatives and it is difficult to represent it graphically on the same scale: we omit its instruction counts in the plot, and its best speed is barely visible.
We see graphically that the benefits of On-Demand vary significantly depending on the task: they are modest for \emph{json2msgpack} and much more significant for \emph{find tweet}. We explain this difference by the fact that in \emph{json2msgpack}, the entire document must be parsed into a data structure, and On-Demand is thus unable to make gains due to its ability to skip some content. On the contrary, in the \emph{find tweet} task, much of the JSON processing can be skipped without adverse consequence. On-Demand still provides a benefit with the \emph{json2msgpack} task because the data is written directly to the desired output without an intermediate storage.

\begin{table}\small
\caption{\label{table:ins}Instructions per input byte for various task and JSON implementations.}
\begin{tabular}{l|ccccc}\toprule
task                &  simdjson (On-Demand) &    simdjson (DOM) & yyjson & RapidJSON &  JSON for Modern C++ \\\midrule
json2msgpack& 4.4 & 5.6 & 6.3 & 24.0 & 374.0 \\
partial tweets& 2.3 & 3.5 & 4.5 & 22.2 & 104. \\
distinct user & 2.2 & 3.4 & 4.5 & 22.1 & 97.1 \\
find tweet& 1.3 & 3.3 & 4.4 & 22.0 & 82.3 \\
top tweet& 2.2 & 3.3 & 4.4 & 22.1 & 115.0 \\
kostya& 3.7 & 5.8 & 6.7 & 18.2 & 105.0 \\
large random& 11.4 & 19.9 & 18.4 & 43.3 & 244.0 \\\midrule
geometric mean& 3.1 & 5.1 & 6.1 & 23.9 & 137.0 \\
\bottomrule
\end{tabular}
\end{table}

\section{Conclusion}

Our work suggests that On-Demand might be a better approach for parsing semi-structured text documents when high efficiency and high expressivity are important. 
To ensure reproducibility, our work is freely available as open source software.

Future work should assess the generality of On-Demand. We should write On-Demand front-end in other programming languages (e.g., Java, Rust, Go). We should apply it to other data sources (e.g., XML, CSV).

There are implementation design strategies that warrant further investigation. For example, our implementation relies  on a precomputed index, giving us the location of structural characters: it makes our implementation convenient and the construction of the index is fast, but an On-Demand front-end may not require an index. Yet an even richer document index might prove more beneficial in some cases: instead of merely indexing the location of the objects, arrays and values, the index could be built with more information about the schema: e.g., the index could inform us that we have encountered an array made of $n$~integers.
Future work should consider heterogeneous computing~\cite{khokhar1993heterogeneous} with On-Demand: the indexing could be computed by a GPU~\cite{10.14778/3377369.3377372} or on an FPGA~\cite{10.1145/3533737.3535094,hahn2022raw,peltenburg2021tens} and passed on to the On-Demand front-end executed on general-purpose processor.

\section*{Acknowledgements}

We thank N.~Boyer for his help writing benchmarks and tests.




\bibliography{ondemand}



\end{document}